\documentclass[aps,twocolumn,showpacs,showkeys,nofootinbib]{revtex4}
\usepackage{epsfig}
\usepackage{amsmath}
\usepackage{amsfonts}
\usepackage{amssymb}
\usepackage{graphicx}
\usepackage{colordvi}
\begin{document}

\title{Weak Field Approach in $f(R)$-Gravity}

\author{$^1${}A. Stabile\footnote{arturo.stabile@gmail.com}, $^2${}S. Capozziello\footnote{capozziello@na.infn.it}}

\affiliation{$^1${}Dipartimento di Ingegneria, Universita' del
Sannio, Palazzo Dell'Aquila Bosco Lucarelli, Corso Garibaldi, 107
- 82100, Benevento, Italy.\\\\
$^2${}Dipartimento di Scienze Fisiche, Universita' di Napoli
\emph{Federico II} and INFN Sez. di Napoli, Compl. Univ. di Monte
S. Angelo, Edificio G, Via Cinthia, I-80126, Napoli, Italy.}

\begin{abstract}

In this communication we discuss the Weak Field Approach, and in
particular the Newtonian limit, applied to $f(R)$-Gravity.
Particular emphasis is placed on the spherically symmetric
solutions and finally, it is clearly shown that General Relativity
results, in the Solar System context, are easily recovered since
Einstein theory is a particular case of $f(R)$-Gravity. This is a
crucial point against several wrong results in literature stating
that these theories are not viable at local scales.

\end{abstract}
 \maketitle

\section{Introduction}

In recent years, the effort to give a physical explanation to the
today observed cosmic acceleration \citep{cosmic_acceleration} has
attracted a good amount of interest in Fourth Order Gravity (FOG)
considered as a viable mechanism to explain the cosmic
acceleration by extending the geometric sector of field equations
without the introduction of Dark Matter and Dark Energy. Other
issues come from Astrophysics. For example, the observed Pioneer
anomaly problem \citep{anderson} can be framed into the same
approach \citep{bertolami} and then a systematic analysis of such
theories urges at small, medium and large scales. Other main topic
is the flatness of the rotation curves of spiral galaxies. In
particular, a delicate point is to address the weak field limit of
any theory of Gravity since two main issues are extremely
relevant: $i)$ preserving the results of General Relativity (GR)
al local scales since they well fit Solar System experiments and
observations; $ii)$ enclosing in a self-consistent and
comprehensive picture phenomena as anomalous acceleration or Dark
Matter at Galactic scales.

The idea to extend Einstein's theory of Gravitation is fruitful
and economic also with respect to several attempts which try to
solve problems by adding new and, most of times, unjustified
ingredients in order to give  self-consistent pictures of
dynamics.
Both the issues could be solved by changing the gravitational
sector, \emph{i.e.} the \emph{l.h.s.} of field equations. In
particular, relaxing the hypothesis that gravitational Lagrangian
has to be only a linear function of the Ricci curvature scalar
$R$, like in the Hilbert-Einstein formulation, one can take into
account an effective action where the gravitational Lagrangian
includes a generic function of Ricci scalar ($f(R)$-Gravity).


In this communication, we report the general approach of the Weak
Field Limit for $f(R)$-Gravity in the metric approach. We deduce
the field equations and derive the weak field potentials with
corrections to the Newtonian potential.

\section{The Field Equations and their Solutions}

Let us start with a general class of $f(R)$-Gravity given by the
action

\begin{eqnarray}\label{HOGaction}
\mathcal{A}\,=\,\int
d^{4}x\sqrt{-g}[f(R)+\mathcal{X}\mathcal{L}_m]
\end{eqnarray}
where $f$ is an unspecified function of curvature invariant $R$.
The term $\mathcal{L}_m$ is the minimally coupled ordinary matter
contribution. In the metric approach, the field equations are
obtained by varying (\ref{HOGaction}) with respect to
$g_{\mu\nu}$. We get

\begin{eqnarray}\label{fieldequationHOG}
f'R_{\mu\nu}-\frac{f}{2}g_{\mu\nu}-f'_{;\mu\nu}+g_{\mu\nu}\Box
f'\,=\,\mathcal{X}\,T_{\mu\nu}
\end{eqnarray}
Here,
$T_{\mu\nu}\,=\,-\frac{1}{\sqrt{-g}}\frac{\delta(\sqrt{-g}\mathcal{L}_m)}{\delta
g^{\mu\nu}}$ is the the energy-momentum tensor of matter, while
$f'\,=\,\frac{df(R)}{dR}$, $\Box\,=\,{{}_{;\sigma}}^{;\sigma}$ and
$\mathcal{X}\,=\,8\pi G$\footnote{Here we use the convention
$c\,=\,1$.}.

The paradigm of Weak Field or Newtonian limit is starting from a
develop of the spherically symmetric metric tensor with respect to
dimensionless quantity $v$. To solve the problem we must start
with the determination of the metric tensor $g_{\mu\nu}$ at any
level of develop (for details see
\citep{newtonian_limit_fR_1,newtonian_limit_fR_2}). From lowest
order of field equations (\ref{fieldequationHOG}) we have
$f(0)\,=\,0$ which trivially follows from the assumption that the
space-time is asymptotically Minkowskian. A such result suggests a
first consideration. If the Lagrangian is developable around a
vanishing value of the Ricci scalar we don't have a cosmological
constant contribution in the $f(R)$-Gravity.

Let us consider a ball-like source with mass $M$ and radius $\xi$.
The energy-momentum tensor $T_{\mu\nu}$ has the components
$T_{tt}\,\sim\,T^{(0)}_{tt}\,=\,\rho$ and
$T_{ij}\,=\,T_{0i}\,=\,0$ where $\rho$ is the mass density (we are
not interesting to the internal structure). The field equations
(\ref{fieldequationHOG}) at $\mathcal{O}(2)$ - order
become\footnote{We set for simplicity $f'(0)\,=\,1$ (otherwise we
have to renormalize the coupling constant $\mathcal{X}$ in the
action (\ref{HOGaction})).}

\begin{eqnarray}\label{PPN-field-equation-general-theory-fR-O2}
\left\{\begin{array}{ll}
R^{(2)}_{tt}-\frac{R^{(2)}}{2}+\frac{\triangle
R^{(2)}}{3m^2}\,=\,\mathcal{X}\,\rho\\\\
\frac{\triangle R^{(2)}}{m^2}-R^{(2)}\,=\,\mathcal{X}\,\rho
\end{array}\right.
\end{eqnarray}
where $\triangle$ is the Laplacian in the flat space,
$R^{(2)}_{tt}$ is the time components of Ricci tensor and
$m^{-2}\,\doteq\,-3f''(0)$. The second line of
(\ref{PPN-field-equation-general-theory-fR-O2}) is the trace of
field equations (\ref{fieldequationHOG}) at $\mathcal{O}(2)$ -
order. It notes that if $f\,\rightarrow\,R$ (\emph{i.e.} $m^2$
diverges) the equations
(\ref{PPN-field-equation-general-theory-fR-O2}) correspond to one
of GR.


The solution for the Ricci scalar $R^{(2)}$ in the third line of
(\ref{PPN-field-equation-general-theory-fR-O2}) is

\begin{eqnarray}\label{scalar_ricci_sol_gen}
R^{(2)}(t,\textbf{x})\,=\,m^2\mathcal{X}\int
d^3\mathbf{x}'\mathcal{G}(\mathbf{x},\mathbf{x}')\rho(t,\mathbf{x}')
\end{eqnarray}
where $\mathcal{G}(\mathbf{x},\mathbf{x}')$ is the Green function
of field operator $\triangle-m^2$. The solution for
$g^{(2)}_{tt}$, from the first line of
(\ref{PPN-field-equation-general-theory-fR-O2}) by considering
that $R^{(2)}_{tt}\,=\,\frac{1}{2}\triangle g^{(2)}_{tt}$, is

\begin{eqnarray}\label{new_sol}
g^{(2)}_{tt}(t,\mathbf{x})\,=\,-\frac{\mathcal{X}}{2\pi}\int
d^3\textbf{x}'\frac{\rho(t,\textbf{x}')}{|\textbf{x}-
\textbf{x}'|}\,\,\,\,\,\,\,\,\,\,\,\,\,\,\,\,\,\,\,\,\,\,\,\,\,\,\,\,
\nonumber\\\\\nonumber\,\,\,\,\,\,\,\,\,\,\,\,\,\,
-\frac{1}{4\pi}\int
d^3\textbf{x}'\frac{R^{(2)}(t,\textbf{x}')}{|\textbf{x}-
\textbf{x}'|}-\frac{2}{3m^2}R^{(2)}(t,\textbf{x})
\end{eqnarray}
We can check immediately that when $f\rightarrow R$ we find
$g^{(2)}_{tt}(t,\textbf{x})\rightarrow-2G\int
d^3\textbf{x}'\frac{\rho(t,\textbf{x}')}{|\textbf{x}-
\textbf{x}'|}$ \citep{postnewtonian_limit_fR}. The solution
(\ref{new_sol}) is the the gravitational potential
$\Phi\,=\,g^{(2)}_{tt}/2$ for $f(R)$-Gravity. We note that $\Phi$
has a Yukawa-like behavior depending by a characteristic length on
which it evolves. As it is evident the Gauss theorem is not valid
since the force law is not $\propto|\mathbf{x}|^{-2}$. The
equivalence between a spherically symmetric distribution and
point-like distribution is not valid and how the matter is
distributed in the space is very important
(\citep{newtonian_limit_R_Ric}).


From the solution (\ref{new_sol}) we can affirm that it is
possible to have solutions non-Ricci-flat in vacuum: \emph{Higher
Order Gravity mimics a matter source}. It is evident from
(\ref{new_sol}) the Ricci scalar is a "matter source" which can
curve the spacetime also in absence of ordinary matter. Besides
the solutions are depending on the only first two derivatives of
$f$ in $R\,=\,0$. So different theories from the third derivative
admit the same solutions.

\section{The Spatial Behaviors of Gravitational Potential}

\begin{figure}[]
\resizebox{\hsize}{!}{\includegraphics[clip=true]{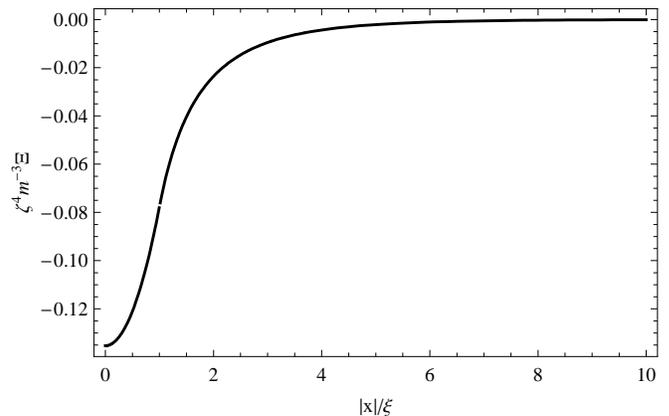}}
\caption{ \footnotesize Plot of dimensionless function
$\zeta^4\mu^{-3}{r_g}^{-1}R^{(2)}$ for
$\zeta\,\doteq\,\mu\xi\,=\,.5$ representing the spatial behavior
of Ricci scalar at second order.} \label{plotricciscalar}
\end{figure}

\begin{figure}[]
\resizebox{\hsize}{!}{\includegraphics[clip=true]{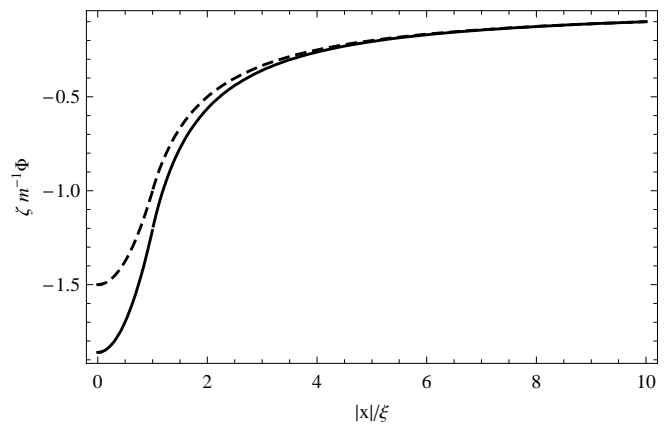}}
\caption{ \footnotesize Plot of metric potential $2\zeta
\mu^{-1}{r_g}^{-1}\Phi$ vs distance from central mass with
$\zeta\,\doteq\,\mu\xi\,=\,.5$. The dashed line is the GR
behavior.} \label{plotpontential00}
\end{figure}

If $m^2\,>\,0$ we have as Green function with spherical symmetry
the following expression

\begin{eqnarray}\label{green_function1}
\mathcal{G}(\mathbf{x},\mathbf{x}')\,=\,-\frac{1}{4\pi}
\frac{e^{-\mu|\mathbf{x}-\mathbf{x}'|}} {|\mathbf{x}-\mathbf{x}'|}
\end{eqnarray}
where we defined $\mu\,\doteq\,\sqrt{|m^2|}$. Then the spatial
behaviors of Ricci scalar (\ref{scalar_ricci_sol_gen}) and
gravitational potential $\Phi$, if $\rho\,=$ constant, are shown
in the Figs. \ref{plotricciscalar} and \ref{plotpontential00}.


For fixed values of the distance $|\mathbf{x}|$, the solution
$g^{(2)}_{tt}$ depends on the value of the radius $\xi$, then the
Gauss theorem does not work also if the Bianchi identities hold.
We can affirm: \emph{the potential does not depend only on the
total mass but also on the mass - distribution in the space}.

It is interesting to note as the gravitational potential assumes
smaller value of its equivalent in GR, then in terms of
gravitational attraction we have a potential well more deep.
Besides if the mass distribution takes a bigger volume, the
potential increases and vice versa.

\begin{figure}[]
\resizebox{\hsize}{!}{\includegraphics[clip=true]{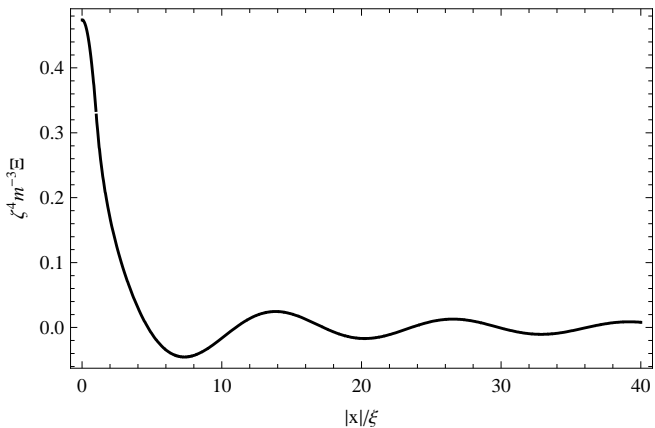}}
\caption{ \footnotesize Plot of dimensionless function
$\zeta^4\mu^{-3}{r_g}^{-1}R^{(2)}$ with
$\zeta\,\doteq\,\mu\xi\,=\,.5$ representing the spatial behavior
of Ricci scalar at second order in the oscillating case.}
\label{plotricciscalar_oscil}
\end{figure}

\begin{figure}[]
\resizebox{\hsize}{!}{\includegraphics[clip=true]{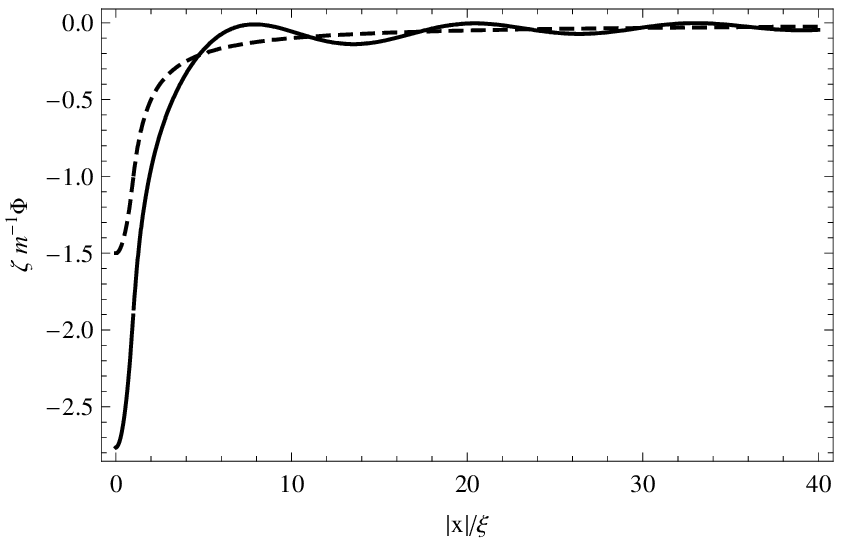}}
\caption{ \footnotesize Plot of metric potential $2\zeta
\mu^{-1}{r_g}^{-1}\Phi$ vs distance from central mass with the
choice $\zeta\,\doteq\,\mu\xi\,=\,.5$ in the oscillating case. The
dashed line is the GR behavior.} \label{plotpontential00_oscil}
\end{figure}

If $m^2\,<\,0$ the Green function assumes the "oscillating"
expression

\begin{eqnarray}\label{green_function_2}
\mathcal{G}(\mathbf{x},\mathbf{x}')\,=\,-\frac{\cos
\mu|\mathbf{x}-\mathbf{x}'|+ \sin
\mu|\mathbf{x}-\mathbf{x}'|}{4\pi\,|\mathbf{x}-\mathbf{x}'|}
\end{eqnarray}
Now the Ricci scalar (\ref{scalar_ricci_sol_gen}) and
gravitational potential $\Phi$ are shown in Figs.
\ref{plotricciscalar_oscil} and \ref{plotpontential00_oscil}.


Finally in the limit of point-like source, \emph{i.e.}
$\rho\,=\,M\,\delta(\mathbf{x})$, we get

\begin{eqnarray}\label{sol_new_pfR}
\left\{\begin{array}{ll}
R^{(2)}\,=\,-r_g\mu^2\frac{e^{-\mu|\mathbf{x}|}}{|\mathbf{x}|}
\\\\
\Phi\,=\,-\frac{r_g}{2}\biggl(\frac{1}{|\textbf{x}|}
+\frac{1}{3}\frac{e^{-\mu|\mathbf{x}|}}{|\mathbf{x}|}\biggr)
\end{array}\right.
\end{eqnarray}
where $r_g\,=\,2GM$ is the Schwarzschild radius. If
$f(R)\,\rightarrow\,R$ we recover the gravitational potential
induced by GR.

\begin{figure}[]
\resizebox{\hsize}{!}{\includegraphics[clip=true]{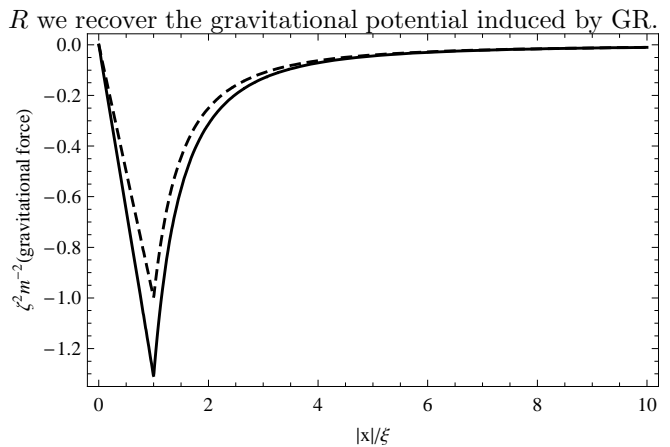}}
\caption{ \footnotesize Comparison between gravitational forces
induced by GR and $f(R)$-Gravity with
$\zeta\,\doteq\,\mu\xi\,=\,.5$.
  The dashed line is the GR behavior.}
\label{plotforce}
\end{figure}

To conclude this section we show in Fig. \ref{plotforce} the
comparison between gravitational forces induced in GR and in
$f(R)$-Gravity in the Newtonian limit. Obviously also about the
force we obtained an intensity stronger than in GR.

\section{Conclusions}

The Weak Field Limit is a crucial issue that has to be addressed
in any relativistic theory of Gravity. It is also the test bed of
such  theories in order to compare them with the well-founded
experimental results of GR, at least at Solar system level.


The general feature  that emerges from the Weak Field Limit is
that correction to the Newtonian potential naturally comes out.
This correction is Yukawa-like term bringing characteristic masse
and length. Conversely, the standard Newtonian potential is just a
feature emerging in the particular case $f(R)\,=\,R$.

It is well-known that the new features related to FOG could have
interesting applications in other fields of Astrophysics as
galactic dynamics, large scale structure and Cosmology in order to
address Dark Matter and Dark Energy issues. The fact that such
"dark" structures have not been definitely discovered at
fundamental quantum scales but operate at large astrophysical
(infra-red scales) could be due to these corrections to the
Newtonian potential which can be hardly detected at laboratory or
Solar System scales.

Finally, the presence of unavoidable light massive modes could
open new opportunities also for the gravitational waves detection
of experiments like VIRGO, LIGO and the forthcoming LISA.


\begin{thebibliography}{99}

\bibitem[Anderson J.D. \emph{et al.} (2002)]{anderson}
Anderson J.D. \emph{et al.} Phys. Rev. D {\bf 65}, 082004 (2002)

\bibitem[Bertolami O. \emph{et all} (2007)]{bertolami}
Bertolami O., B\"{o}hmer C.G., Harko T., Lobo F.S.N., Phys. Rev. D
\textbf{75}, 104016 (2007)

\bibitem[Capozziello S. \emph{et all} (2007)]{newtonian_limit_fR_1}
Capozziello S., Stabile A., Troisi A., Phys. Rev. D {\bf 76},
104019 (2007)

\bibitem[Capozziello S. \emph{et all} (2009)]{newtonian_limit_fR_2}
Capozziello S., Stabile A., Troisi A., Modern physics letters A
{\bf 24}, 659 (2009)

\bibitem[Capozziello S. \emph{et all} (2009)]{newtonian_limit_R_Ric}
Capozziello S., Stabile A., Class. Quant. Grav. {\bf 26}, 085019
(2009) {\bf 24}, 659 (2009)

\bibitem[Perlmutter S. \emph{et al.} (2006)]{cosmic_acceleration}
Perlmutter S. \emph{et al.} Astron. Astrophys. {\bf 447}, 31
(2006)

\bibitem[Stabile A. (2009)]{postnewtonian_limit_fR}
Stabile A., accepted by Phys. Rev. D (arXiv: gr-qc/1004.1973v2)

\end{thebibliography}
\end{document}